\documentstyle[twocolumn,epsf,amsfonts,aps,pre]{revtex}   

\begin{document}

\draft

\twocolumn[\hsize\textwidth\columnwidth\hsize\csname 
@twocolumnfalse\endcsname                            

\preprint{\sf Manuscript version of July 24, 2000}

%
%

\title{Towards Better Integrators for Dissipative Particle Dynamics 
       Simulations}

\author{Gerhard Besold$^{1}$\cite{corresp}, Ilpo Vattulainen$^{1}$, 
        Mikko Karttunen$^{2}$, and James M. Polson$^{3}$} 
        
\address{
$^1$Department of Chemistry, Technical University of Denmark, 
        Building 207, DK--2800 Lyngby, Denmark\\
$^2$Max Planck Institute for Polymer Research, Theory Group,
        PO Box 3148, D--55021 Mainz, Germany\\
$^3$Department of Physics, University of Prince Edward Island,
        550 University Avenue, Charlottetown, PEI, Canada C1A 4P3
}

\date{Accepted for publication in PHYSICAL REVIEW E as Rapid Communication
--- tentative publication issue: 01 Dec 2000}

\maketitle

\begin{abstract}

\hspace*{3mm}Coarse--grained models that preserve hydrodynamics provide a natural 
approach to study collective properties of soft--matter systems.
Here, we demonstrate that commonly used integration schemes in 
dissipative particle dynamics give rise to pronounced artifacts in 
physical quantities such as the compressibility and the diffusion 
coefficient. We assess the quality of these integration schemes, 
including variants based on a recently suggested self--consistent 
approach, and examine their relative performance. Implications of 
integrator--induced effects are discussed.

\end{abstract}

\pacs{PACS number(s):\, 02.70.Ns, 47.11.+j, 05.40.--a} 



\vskip2pc]                                           

%
One of the current challenges in theoretical physics is to understand the basic 
principles that govern collective properties of soft--matter systems. 
From a modeling point of view, these systems are problematic due to 
the fact that numerous phenomena take place at mesoscopic time and 
length scales, while the most accurate ``brute--force'' molecular 
dynamics simulations are limited to microscopic time and length scales. 
To overcome this problem, a number of ``coarse--grained'' approaches
\cite{Hoo92,Esp95a,coarse_grained_methods} have been suggested and 
developed to simplify the underlying microscopic model while retaining 
the essential physics.

%
Introduced in 1992 \cite{Hoo92} and cast into its present form in 1995 
\cite{Esp95a}, Dissipative Particle Dynamics (DPD) has become one of 
the most promising methods for soft--matter simulations \cite{Gro97,War98}.
From a technical point of view, DPD differs from Molecular Dynamics 
(MD) in two respects. First, the conservative pairwise forces between 
DPD particles (which represent clusters of microscopic particles) are 
soft--repulsive, which makes it possible to extend the simulations to 
longer time scales. Second, a special ``DPD thermostat'' for the 
canonical ensemble is implemented in terms of dissipative as well as 
random pairwise forces such that the momentum is locally conserved, 
which results in the emergence of hydrodynamic flow effects on the 
macroscopic scale.

%
However, the  pairwise coupling of particles by the dissipative and 
random forces makes the integration of the equations of motion a 
non--trivial task. It has been observed that essentially all traditional 
integration schemes lead to distinct deviations from the true equilibrium 
behavior, including an unphysical systematic drift of the temperature from 
the value predicted by the fluctuation--dissipation theorem, and 
artificial structures in the radial distribution function 
\cite{Gro97,Nov98,Mar97a,Pag98,Gib99,Mar97c}. Consequently, various 
integration schemes have been suggested to overcome these problems 
\cite{Gro97,Pag98,Gib99}. Some approaches are based on the use of 
phenomenological ``tuning parameters'' which mimic higher--order 
corrections in the integration procedure \cite{Gro97,Gib99}. A more 
elaborate technique suggested by Pagonabarraga {\it et al.}\/ 
\cite{Pag98} determines the velocities and velocity--dependent 
dissipative forces in a self--consistent fashion. Although both 
approaches have been shown to reduce numerical artifacts in some cases 
\cite{Gro97,Nov98,Pag98,Gib99}, there is still no good understanding 
as to which integration scheme is most suitable for future extensive 
soft--matter DPD simulations, and a thorough comparison including 
recently suggested schemes \cite{Pag98,Gib99} is pending. Moreover, 
the effect of integrators on dynamic quantities such as transport 
coefficients has received only little attention so far \cite{transp}.
In this work, we address these issues.

%
We consider various 
integrators based on the velocity--Verlet scheme \cite{Ver67} and assess 
their quality by studying a number of physical observables such as 
temperature, radial distribution function, compressibility, and tracer 
diffusion. We demonstrate that there is no reason to use integrators 
which contain tuning parameters, since better schemes are readily 
available. However, we also find that even a self--consistent approach 
gives rise to subtle temperature drifts, which can be corrected by a 
method presented in this work.

%
For a system of $N$ particles with mass $m$, coordinates $\{{\bf r}_i\}$, 
and velocities $\{{\bf v}_i\}$, the pairwise conservative, dissipative, 
and random forces exerted on particle ``$i$'' by particle ``$j$'' are 
given by, respectively,
\begin{mathletters}
\label{DPD_forces}
\begin{eqnarray}
{\bf F}_{ij}^C &=& \alpha\ \omega (r_{ij})\  
   {\bf e}_{ij}\,,\label{forces:1}\\
{\bf F}_{ij}^D &=& -\gamma\ \omega^2 (r_{ij})\
   ({\bf v}_{ij}\!\cdot\! {\bf e}_{ij})\, {\bf e}_{ij}\,,\\
{\bf F}_{ij}^R &=& \sigma\ \omega (r_{ij})\ \xi_{ij}\ 
   {\bf e}_{ij}\,,\label{forces:2}
\end{eqnarray}
\end{mathletters}
where ${\bf r}_{ij} = {\bf r}_i - {\bf r}_j$, $r_{ij} = | {\bf r}_{ij} |$,
${\bf e}_{ij} = {\bf r}_{ij} / r_{ij}$, and 
${\bf v}_{ij} = {\bf v}_i - {\bf v}_j$. 
The $\xi_{ij}$ are symmetric random variables with zero mean and unit 
variance, uncorrelated for different pairs of particles and different 
times. The forces are soft--repulsive due to the weight function 
\,$\omega(r_{ij})$\, for which we adopt the commonly made choice 
$\omega(r_{ij}) = 1 - r_{ij}/r_c$\, for \,$r_{ij}\!\le\!r_c$,\, 
and\, $\omega(r_{ij})=0$\, for \,$r_{ij}\!>\!r_c$\,, with a cut--off 
distance \,$r_c$\, \cite{Gro97}. The strength of the conservative,
dissipative, and random forces is determined by the parameters $\alpha$, 
$\gamma$, and $\sigma$, respectively. The equations of motion are then 
given by the set of stochastic differential equations 
\vspace*{-2mm}
\begin{mathletters} 
\label{equs_of_motion}
\begin{eqnarray} 
d {\bf r}_i &=& {\bf v}_i\,dt\,,\label{motion:1}\\
d {\bf v}_i &=& \frac{\displaystyle 1}{\displaystyle m}
   \left( {\bf F}_i^{C}\,dt + {\bf F}_i^{D}\,dt + 
   {\bf F}_i^R\,\sqrt{dt}\,\right)\,,\label{motion:2}
\label{equs_of_motion2}
\end{eqnarray}
\end{mathletters}
where \,${\bf F}_i^{C}=\sum_{j \not= i}{\bf F}_{ij}^{C}$\, is the total 
conservative force acting on particle ``$i$'' (with ${\bf F}_i^{D}$ and 
${\bf F}_i^{R}$ defined correspondingly). This continuous--time version 
of DPD satisfies detailed balance and describes the canonical ensemble
if  \,$\sigma$\, and \,$\gamma$\, obey the fluctuation--dissipation 
relation \,$\sigma^2/\gamma=2\,k_B T^{\ast}$ \cite{Esp95a}.

%
In order to integrate the equations of motion, we use the widely adopted 
velocity--Verlet scheme \cite{Ver67} as a starting point and consider the 
most commonly used integrators based on this approach. These are summarized 
in Table~\ref{table1}, where the acronym ``MD--VV'' corresponds to the 
standard velocity--Verlet algorithm used in classical MD simulations. 
Unlike in MD, however, the forces in DPD depend on the velocities. For that 
reason Groot and Warren \cite{Gro97} proposed a modified velocity--Verlet 
integrator (``GW($\lambda$)'' in Table~\ref{table1}). In this approach, 
the forces are still updated only once per integration step, but the 
dissipative forces are evaluated based on intermediate ``predicted'' 
velocities $\widetilde{\bf v}_{i}$. The calculation of 
$\widetilde{\bf v}_i$ involves the use of a phenomenological tuning 
parameter $\lambda$ which mimicks higher--order corrections in the 
integration procedure. The problem is that the optimal value of $\lambda$, 
which minimizes temperature drift and other artifacts, depends on model 
parameters and has to be determined empirically. Recently, Gibson 
{\it et al.} \cite{Gib99} proposed a slightly modified version of the 
GW integrator. \,This \,``GCC($\lambda$)''
%
%
\narrowtext
\begin{table}[!]
\caption{Update schemes for a single integration step
for various DPD integrators (acronyms see text).\\
\hspace*{2em}GW($\lambda$)\,:\hfill steps (0)--(4),\,(s)\hspace*{2.8em}\\
\hspace*{2em}MD--VV $\equiv$ GW($\lambda\!=\! 1/2 $)\,:\hfill steps
(1)--(4),\,(s)\
   ${}^{\rm a}$\hspace*{2em}\\
\hspace*{2em}GCC($\lambda$)\,:\hfill steps (0)--(5),\,(s)\hspace*{2.8em}\\
\hspace*{2em}DPD--VV $\equiv$ GCC($\lambda\!=\! 1/2 $)\,:\hfill steps
(1)--(5),\,(s)\
   ${}^{\rm a}$\hspace*{2em}
}
\label{table1}
\begin{tabular}{ll}\\[-2mm]
(0) & ${\widetilde{\bf v}}_i\ \longleftarrow\ {\bf v}_i\ +\
{\displaystyle \lambda\,\frac{1}{m}} \left( {\bf F}_i^C\,\Delta t +
  {\bf F}_i^D\,\Delta t +
{\bf F}_i^R\,\sqrt{\Delta t}\,\right)$\\[3mm]
(1) & ${\bf v}_i\ \longleftarrow\ {\bf v}_i\ +\
{\displaystyle \frac{1}{2}\,\frac{1}{m}} \left( {\bf F}_i^C\,\Delta t +
  {\bf F}_i^D\,\Delta t +
{\bf F}_i^R\,\sqrt{\Delta t}\,\right)$\\[4mm]
(2) & ${\bf r}_i\,\ \longleftarrow\ {\bf r}_i\hspace*{0.5mm}\ +\
{\bf v}_i\,\Delta t$\\[4mm]
(3) & Calculate\quad ${\bf F}_i^C\{{\bf r}_j\},
\quad {\bf F}_i^D\{{\bf r}_j,\,{\widetilde{\bf v}}_j\},\quad
  {\bf F}_i^R\{{\bf r}_j\}$\\[3mm]
(4) & ${\bf v}_i\ \longleftarrow\ {\bf v}_i\ +\
{\displaystyle \frac{1}{2}\,\frac{1}{m}} \left( {\bf F}_i^C\,\Delta t +
  {\bf F}_i^D\,\Delta t +
{\bf F}_i^R\,\sqrt{\Delta t}\,\right)$\\[4mm]
(5) & Calculate\quad ${{\bf F}_i^D}\{{\bf r}_j,\,
{{\bf v}_j}\}$\\[1mm]
(s)\tablenotemark[2] & Calculate\quad ${k_B T} \!= {\displaystyle
  \frac{m}{3N\!-\!3}}\,
\displaystyle{\sum\limits_{i=1}^N}
\ {\bf v}_i^{\,2}$\,,\quad\ldots\\[4mm]
\end{tabular}
\tablenotetext[1]{
with substitution of \,${\bf v}_j$\, for \,${\widetilde{\bf v}}_j$\,
in step (3).
}
\tablenotetext[2]{
Sampling step (calculation of temperature $k_B T$\,, $g(r)$, \ldots)
}
\end{table}
\noindent
integrator updates the dissipative 
forces (step (5) in Table~\ref{table1}) for a second time at the end of 
each integration step. Choosing  $\lambda\!=\! 1/2$  in the GCC integrator 
is equivalent to the MD--VV scheme supplemented by the second update of 
the dissipative forces. This Verlet--type integrator, here termed 
``DPD--VV'', is appealing because it does {\it not}\/ involve a tuning 
parameter, yet takes the velocity--dependence of the dissipative forces 
at least approximately into account.

\begin{figure}[!]
\centerline{
  \epsfxsize=80mm
  \epsfbox{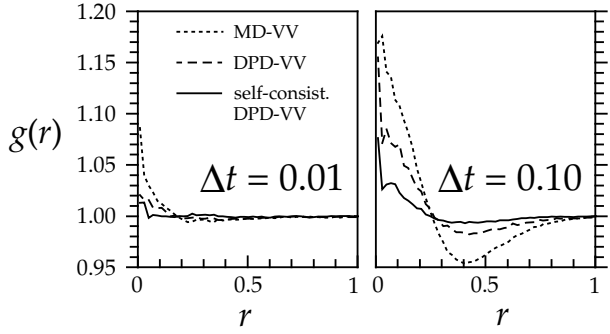}
}
\vspace*{0.2cm}
\caption{
Radial distribution functions $g(r)$ as obtained in DPD simulations
of the 3D ideal gas for $\Delta t=0.01$ (left) and $\Delta t=0.1$
(right) for velocity--Verlet--based integrators. Time is given in
units of $r_c \sqrt{m/k_B T}$.
}
\label{fig1}
\end{figure}

%
Unfortunately, all of the above integrators display pronounced unphysical 
artifacts in $g(r)$ and thus do not produce the correct equilibrium 
properties (see Fig.~\ref{fig1} and discussion below). This highlights 
the need for an approach in which the velocities and dissipative forces 
are determined in a self--consistent fashion. To this end, 
we present in Table~\ref{table2} the update schemes for two 
self--consistent variants of DPD--VV. The basic variant, which is similar 
in spirit to the self--consistent leap--frog scheme introduced by 
Pagonabarraga {\it et al.}\/ \cite{Pag98}, determines the velocities and 
dissipative forces self--consistently through functional iteration, and 
the convergence of the iteration process is monitored by the instantaneous 
temperature $k_B T$. In the second approach, we furthermore couple the 
system to an auxiliary thermostat, thus obtaining an ``extended--system'' 
method in the spirit of Nos{\'e}--Hoover \cite{Thi99} (see below for 
details).

%
Since the problems due to dissipative and stochastic forces in DPD are 
particularly noticeable in the absence of \,\,conservative forces, \,\,we 
\,\,therefore \,\,focus \,\,on \,\,a \,\,3D \,ideal \,gas \cite{Mar97a,Pag98}. 
\,In \,our \,simulations \,we \,use \,a \,box \,of \,size
%
%
\narrowtext
\begin{table}[!]
\caption{Update scheme
for self-consistent DPD--VV without and with
(steps (i)--(iii)) auxiliary thermostat. The self--consistency
loop is over steps (4b) and (5) as indicated.
The desired temperature is $k_B T^{\ast}$.
Initialization: $\eta=0$, $\gamma=\sigma^2/(2 k_BT^{\ast})$, 
and $k_B T$ calculated from the initial velocity distribution.
}
\label{table2}
\begin{tabular}{lll}\\[-2mm]
{} & (i) & ${\bf \dot{\eta}}\ \hspace*{0.47em}\longleftarrow\ C\,(k_B T -
   k_B T^{\displaystyle \ast})$\\[3mm]
{} & (ii) & ${\eta}\ \hspace*{0.47em}\longleftarrow\ \eta\ +\ {\bf \dot{\eta}}\,
   \Delta t$\\[2mm]
{} & (iii) & $\gamma\ \hspace*{0.37em}\longleftarrow\ {\displaystyle
   \frac{\sigma^2}{2\,k_B T^{\displaystyle \ast}}}\,(1 +
   \eta\,\Delta t)$\\[3mm]
{} & (1) & ${\bf v}_i\ \longleftarrow\ {\bf v}_i\ +\
   {\displaystyle \frac{1}{2}\,\frac{1}{m}} \left( {\bf F}_i^C\,\Delta t +
   {\bf F}_i^D\,\Delta t + {\bf F}_i^R\,\sqrt{\Delta t}\,\right)$\\[4mm]
{} & (2) & ${\bf r}_i\,\ \longleftarrow\ {\bf r}_i\hspace*{0.5mm}\ +\
   {\bf v}_i\,\Delta t$\\[4mm]
{} & (3) & Calculate\quad ${\bf F}_i^C\{{\bf r}_j\}, \quad {\bf F}_i^D
   \{{\bf r}_j,\,{\bf v}_j\},\quad {\bf F}_i^R\{{\bf r}_j\}$\\[3mm]
{} & (4a) & $\widehat{{\bf v}}_i\ \longleftarrow\ {\bf v}_i\ +\
   {\displaystyle \frac{1}{2}\,\frac{1}{m}} \left\{ {\bf F}_i^C\,\Delta t +
   {\bf F}_i^R\,\sqrt{\Delta t}\,\right\}$\\[3mm]
   {} & (4b) & ${\bf v}_i\ \longleftarrow\ \widehat{{\bf v}}_i\ +\
   {\displaystyle \frac{1}{2}\,\,\frac{1}{m}}\,
   {\bf F}_i^D\,\Delta t$\\[-4mm]
\setlength{\unitlength}{0.1em}
\begin{picture}(6,30)
 \put(0,3){\line(1,0){6}}
 \put(0,3){\line(0,1){23}}
 \put(0,26){\vector(1,0){6}}
\end{picture}&
(5) & Calculate\quad ${{\bf F}_i^D}\{{\bf r}_j,\,
{{\bf v}_j}\}$\\[1mm]
{} & (s) & Calculate\quad ${k_B T} \!= {\displaystyle \frac{m}{3N\!-\!3}}\,
\displaystyle{\sum\limits_{i=1}^N}
\ {\bf v}_i^{\,2}$\,,\quad\ldots\\[4mm]
\end{tabular}
\end{table}
%
%
\noindent
$10\!\times\!10\!\times\!10$ with periodic boundary 
conditions, a random force strength $\sigma = 3$\,, and a particle density 
$\rho = 4$\, (i.e., $N = 4000$ particles) \cite{units}. For the 
equilibrated system, the temperature $\langle k_B T \rangle$ and the 
radial distribution function \,$g(r)$\, were sampled. For the ideal gas,  
$g(r)\!\equiv\!1$ in the continuum limit, and therefore any deviation 
from 1 has to be interpreted as an artifact due to the employed 
integration scheme. Artifacts in $g(r)$ are also reflected in the relative 
isothermal compressibility 
\,${\widetilde\kappa}_T\equiv\kappa_T/\kappa_T^{\scriptstyle {\em ideal}}$, 
where \,$\kappa_T^{\scriptstyle {\em ideal}} = (\rho\,k_B T^{\ast})^{-1}$\,  
denotes the compressibility of the ideal gas in the continuum limit. For 
an arbitrary fluid, ${\widetilde \kappa}_T$ is related to $g(r)$ by 
\,${\widetilde \kappa}_T = 1 + 4\pi\rho\!\int_0^{\infty}\!dr\,\,r^2
\,[g(r) - 1]$\,, and thus any deviation from \,${\widetilde \kappa}_T\!=
\!1$\, for the ideal gas indicates an integrator--induced artifact. 
Finally, to gauge underlying problems in the actual {\it dynamics} of the 
system, we consider the tracer diffusion coefficient $D_T = \lim_{t \to 
\infty} \frac{1}{6Nt} \sum_{i=1}^{N} \langle [ {\bf r}_i(t) - 
{\bf r}_i(0) ]^2 \rangle $, which characterizes temporal correlations 
between the displacements (velocities) of the tagged particle.

%
Results for $g(r)$ are shown in Fig.~\ref{fig1}. We find that the 
deviations from $g(r) = 1$ are very pronounced for MD--VV, indicating 
that even at small time steps it gives rise to unphysical correlations. 
The performance of DPD--VV is clearly better, while the self--consistent 
scheme leads to even smaller deviations. Studies of the integrators 
GW($\lambda=0.65)$ \,and \,GCC($\lambda$) \,(for \,a \,few \,values 
\,of \,$\lambda$) re-
\vspace*{1mm}
%
\begin{figure}[!]
\centerline{
  \epsfxsize=80mm
  \epsfbox{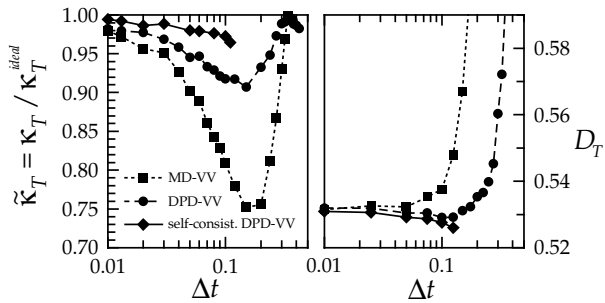}
}
\vspace*{0.2cm}
\caption{
Left: Dimensionless isothermal compressibility ${\widetilde \kappa}_T$
vs. $\Delta t$ for the integrators shown in the legend.
Right: Tracer diffusion coefficient $D_T$
vs. $\Delta t$ for the same integrators.
}
\label{fig2}
\end{figure}
%
%
%
\begin{figure}[!]
\centerline{
  \epsfxsize=60mm
  \epsfbox{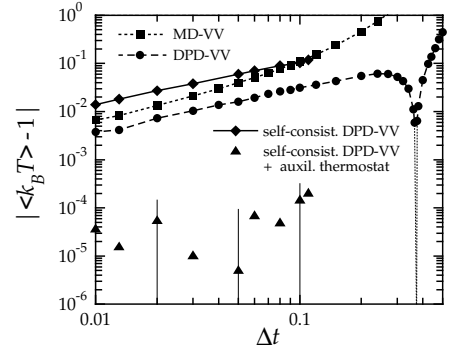}
}
\vspace*{0.2cm}
\caption{
Double--logarithmic plot of the modulus of the deviation
of $<\!\!k_B T\!\!>$ from the desired temperature
\,$k_B T^{\ast}\!\equiv\!1$ vs. $\Delta t$. For
self--consistent DPD--VV with auxiliary thermostat,
$1\sigma$ error bars are shown for some of the data points.
}
\label{fig3}
\end{figure}
\noindent
vealed that their results were approximately similar to those of MD--VV 
and DPD--VV, respectively. For all integrators, the artificial structure 
in $g(r)$ becomes more pronounced with increasing time increment 
$\Delta t$, and it is intriguing that the bias introduced by the 
self--consistent integrator for $\Delta t\!=\!0.10$ is comparable to that 
introduced by MD--VV for $\Delta t\!=\!0.01$.

%
The relative isothermal compressibilities $\widetilde{\kappa}_T$ 
evaluated from $g(r)$ are shown in Fig.~\ref{fig2}. The qualitative 
behavior of $\widetilde{\kappa}_T$ reflects our findings for $g(r)$ 
\cite{compr}. However, the magnitude of deviations from 
$\widetilde{\kappa}_T = 1 $ is astounding, and raises serious concern 
for studies of response functions such as the compressibility for 
interacting fluids close to phase boundaries. Similarly, the results for 
tracer diffusion (also in Fig.~\ref{fig2}) indicate that DPD--VV and the 
self--consistent approach work well up to reasonably large time steps, 
while the other integrators were found to perform less well. Thus, the 
decay of velocity correlations in tracer diffusion is sensitive to the 
choice of the integrator. These results demonstrate that special care is 
needed in studies of DPD model systems, and suggest that integrators 
commonly used in MD should not be employed in DPD as such.

%
Next we discuss the deviations of the observed actual temperature 
$\langle k_B T\rangle$ from the desired temperature $k_B T^{\ast}$ 
(see Fig.~\ref{fig3}). For MD--VV this ``temperature drift'' is always 
positive and increases monotonically  with $\Delta t$. For DPD--VV,  
$\langle k_B T\rangle$ first decreases with increasing $\Delta t$, 
then exhibits a minimum at $\Delta t\!\approx\!0.25$, and eventually 
becomes larger than $k_B T^{\ast}$. The self--consistent approach 
exhibits a negative, monotonically increasing temperature drift up to 
$\Delta t\!\approx\!0.13$, where this scheme becomes unstable at the 
employed particle density. Most importantly, we find, surprisingly, that 
the modulus of the temperature deviation is even larger than the one for 
DPD--VV. In a recent work, Pagonabarraga {\it et al.} \cite{Pag98} studied the 
2D ideal gas using a self--consistent version of the leap--frog algorithm, 
and found good temperature control for $\Delta t = 0.06$ at $\rho = 0.5$. 
This discrepancy can be explained by our observation for the 3D ideal gas
that the temperature drift is in general more pronounced at higher 
densities.

%
In cases where temperature preservation is crucial in calculating 
equilibrium quantities, we finally demonstrate how this can be achieved. 
The idea is to supplement the self--consistent scheme by an auxiliary 
thermostat, which preserves the pairwise conservation of momentum by 
employing a {\it fluctuating}\/ dissipation strength 
\begin{equation} 
\label{thermostat}
\gamma(t) = \frac{\sigma^2}{2\,k_B T^{\ast}}\,( 1 + \eta(t)\,\Delta t)\,,
\end{equation} 
where \,$\eta$\, is a thermostat variable. The rate of change of $\eta$ 
is proportional to the instantaneous temperature deviation, 
\,${\bf \dot{\eta}}=C(k_B T\!-\!k_B T^{\ast})$\,, where $C$ is a coupling 
constant (step (i) in Table~\ref{table2}) \cite{constant_c}. This 
first--order differential equation must be integrated (step (ii)) 
simultaneously with the equations of motion. In this respect our 
thermostat resembles the Nos\'e{}--Hoover thermostat familiar from MD 
simulations \cite{Thi99}.

%
For this extended--system method, we find (Fig.~\ref{fig3}) that the 
temperature deviations diminish by over two orders of magnitude, with a 
modulus typically of the order of $10^{-5} \ldots 10^{-4}$. We also found 
virtually the same results for $g(r)$ and $\widetilde{\kappa}_T$ as for 
the self--consistent scheme without the thermostat. This suggests that the 
auxiliary thermostat is useful in studies of equilibrium quantities such 
as the speficic heat. However, we feel that the auxiliary thermostat is 
not an ideal approach to describe quantitative aspects of tracer diffusion. 
A more detailed study is currently in progress \cite{interactive}.

%
In this work, we have shown that integration schemes may in DPD lead to 
pronounced artifacts in response functions and transport coefficients. 
This constitutes a serious problem for studies of soft systems, and 
highlights the timely need to resolve this issue. We have demonstrated 
that these artifacts can be sufficiently suppressed by using 
velocity--Verlet--based schemes in which the velocity dependence of the 
dissipative forces is taken into account. The velocity--Verlet scheme 
without iterations but with an additional update of the dissipative forces
(DPD--VV) performs --- at essentially unchanged computational costs ---
already considerably better than the Groot--Warren integrator. The best 
overall performance is found for a recently proposed approach \cite{Pag98}, 
in which particle velocities and velocity--dependent dissipative forces 
are determined self--consistently. This scheme provides an accurate 
description for the quantities studied here, except for the temperature 
whose persisting drift is found to be unexpectedly significant. As shown 
in this work, however, this drift can be suppressed by at least two orders 
of magnitude by a scheme which supplements self--consistency with an 
auxiliary thermostat. The computational cost of the self--consistent 
scheme without thermostat remains modest, and increases only by a factor 
of 1.5 to 3 with respect to the standard velocity--Verlet algorithm.

%
Although DPD has been very successful especially in simulations of 
polymeric systems and in reproducing equilibrium properties, there have 
been doubts as to whether DPD is able to describe the dynamics and 
transport properties of complex fluids \cite{Gro97,Pag98}. Since 
the origin of the observed discrepancies is not clear and a general 
theory is still lacking, it will be interesting to see what the 
impact of the ideas presented here is on transport properties. 
Work is in progress to address these questions \cite{interactive}. 

%
This work was supported by the Danish Natural Science Research Council 
(G.B.), a European Union grant (I.V.), and the Natural Sciences and 
Engineering Research Council of Canada (J.M.P.).

\vspace*{-2mm}

%
%


\begin{thebibliography}{100}
\bibitem{units}
As basic units we use the particle mass $m$, the cut--off length 
$r_c$, and the desired equilibrium temperature $k_B T^{\ast}$.
\bibitem{constant_c} 
The coupling constant $C$ has to be chosen with care. 
For the simulations reported here, we always ensured that the 
employed values of $C$ did not cause any bias.
\end{thebibliography}
\end{document}